\documentclass[a4paper,11pt]{article}
\usepackage{aaskaiid}
\usepackage{orcidlink}
\usepackage[T1]{fontenc}
\usepackage{amsfonts, amsmath}
\usepackage{array}
\usepackage{multirow}
\usepackage{float}

\newcommand{\mbh}{M_{\bullet}}

\def\gs{\mathrel{\lower0.6ex\hbox{$\buildrel {\textstyle >}
 \over {\scriptstyle \sim}$}}}
\def\ls{\mathrel{\lower0.6ex\hbox{$\buildrel {\textstyle <}
 \over {\scriptstyle \sim}$}}}

\def\lta{\mathrel{\spose{\lower 3pt\hbox{$\mathchar"218$}}
     \raise 2.0pt\hbox{$\mathchar"13C$}}}
\def\gta{\mathrel{\spose{\lower 3pt\hbox{$\mathchar"218$}}
     \raise 2.0pt\hbox{$\mathchar"13E$}}}
\def\HI{\hbox{H~$\scriptstyle\rm I\ $}}
\def\HII{\hbox{H~$\scriptstyle\rm II\ $}}

\def\kms{\,{\rm km}\,{\rm s}^{-1}}

\def\and{{\rm M31}}

\def \kms{\;{\rm km}\,{\rm s}^{-1}}

\def\oii{\mbox{[O\,{\sc ii]\sc{$\lambda$3727}}}}

\def\oiii{\mbox{[O\,{\sc iii]\sc{$\lambda$5007}}}}
\def\nii{\mbox{[N\,{\sc ii]\sc{$\lambda$6585}}}}

\def\sii{\mbox{[S\,{\sc ii]\sc{$\lambda$6718,6732}}}}
\def\halpha{\mbox{H\,{\sc $\alpha$}}}
\def\hbeta{\mbox{H\,{\sc $\beta$}}}

\title{Unlocking the Full Potential of SKAO Extra-galactic Science with High-multiplex Optical Spectroscopy}
\ShortTitle{Unlocking the Full Potential of the SKAO with Spectroscopy}

\author[1]{Kenneth J Duncan\orcidlink{0000-0001-6889-8388}}
\ShortName{Kenneth J Duncan et al.} 
\author[2]{Natasha Maddox\orcidlink{0000-0001-8312-5260}}
\author[3]{Daniel J. B Smith\orcidlink{0000-0001-9708-253X}}
\author[1]{Marina I Arnaudova\orcidlink{0000-0002-1128-0592}}
\author[1,4]{Catherine L Hale\orcidlink{0000-0001-6279-4772}}
\author[1]{Sophia Flury\orcidlink{0000-0002-0159-2613}}
\author[3]{Luke Holden\orcidlink{0000-0002-1721-1918}}
\author[4]{Matt J Jarvis\orcidlink{0000-0001-7039-9078}}
\author[6,7]{Elizabeth A. K. Adams\orcidlink{0000-0002-9798-5111}}
\author[5]{S. Ilani Loubser\orcidlink{0000-0002-3937-7126}}
\author[8]{Mamta Pandey-Pommier\orcidlink{0000-0001-5829-1099}}
\author[3,4]{Anastasia Ponomareva\orcidlink{0000-0003-4100-0173}}

\affiliation[1]{Institute for Astronomy, University of Edinburgh, Blackford Hill, Edinburgh, EH9 3HJ, UK}
\emailAdd{kdun@roe.ac.uk}
\affiliation[2]{School of Physics, H.H. Wills Physics Laboratory, Tyndall Avenue, University of Bristol, Bristol, BS8 1TL, UK}
\affiliation[3]{Centre for Astrophysics Research, School of Physics, Astronomy and Mathematics, University of Hertfordshire, College Lane, Hatfield AL10 9AB, UK}
\affiliation[4]{Astrophysics, Denys Wilkinson Building, Department of Physics, University of Oxford, Keble Road, Oxford OX1 3RH, UK}
\affiliation[5]{Centre for Space Research, North-West University, Potchefstroom 2520, South Africa}
\affiliation[6]{ASTRON, the Netherlands Institute for Radio Astronomy, Oude Hoogeveenseweg 4, 7991 PD Dwingeloo, The Netherlands}
\affiliation[7]{Kapteyn Astronomical Institute, University of Groningen, P.O. Box 800, 9700 AV, Groningen, The Netherlands}
\affiliation[8]{University Catholic of Lyon, 10, place des Archives 69288, Lyon Cedex 02, France}

\abstract{Parallel to the transformation in radio continuum and \HI observations from SKAO pathfinders and precursors over the past decade has been the development of a new generation of multi-object spectrographs that will be equally transformational in a broad range of scientific areas. For all extragalactic radio source populations, this spectroscopy is essential for providing precise redshifts, separating star-formation and AGN activity, identifying accretion modes and revealing detailed host galaxy properties. 
It is only with the detailed emission line and optical continuum diagnostics from spectroscopy that we can begin to link the AGN and star-formation activity revealed by the radio continuum to the kinematic and chemical histories of galaxies.
Crucially, extensive spectroscopy also unlocks the full potential of \HI observations by enabling statistical measures of the \HI content of galaxies out to $z=1$ and beyond, through spectral stacking analyses, as well as comprehensively tracing the local environment and large-scale structure to enable studies of environmental effects on the baryon cycle.
We outline the scientific synergies enabled by combining SKAO continuum and \HI surveys with current optical spectroscopic surveys, as well as identifying the needs and scientific potential of future spectroscopic surveys dedicated to the radio source population with both existing and planned optical facilities.}

\begin{document}
\maketitle

\section{Introduction}

The SKA Observatory is expected to be a transformational facility for observing the radio continuum and spectral Universe \citep{braun2019anticipatedperformancesquarekilometre}, extending these observables into new regimes of sensitivity, resolution (spatial and spectral) and survey efficiency. However, observational astronomy is an increasingly multi-wavelength endeavour. As is the case for current and past radio facilities, achieving the full scientific potential of SKAO continuum and neutral hydrogen (\HI) extragalactic surveys will require combining these radio observations with a host of complementary observations across the electromagnetic spectrum to provide a comprehensive picture of galaxy properties.

Historically, one of the key synergies has been the targeted follow-up of radio-continuum-selected sources with optical spectroscopy, extending back to the milestone discovery of quasi-stellar objects \citep[QSO;][]{schmidt1963_qso} and the enormous cosmological and astrophysical implications this revealed.
Since the early systematic follow-up of the brightest radio sources in the sky \citep{smith1976_3c, kristian1978_3c}, the size and scope of both radio and optical spectroscopy surveys has evolved enormously. 
Most notably, the large homogeneous samples of spectra provided by the Sloan Digital Sky Survey \citep[SDSS;][]{york2000_sdss, strauss2002_sdssmgs, richards2002_sdssqso} have been transformative across many areas of galaxy and active galactic nuclei (AGN) evolution. 

When combined with wide area radio continuum surveys such as the Faint Images of the Radio Sky at Twenty Centimeters \citep[FIRST;][]{becker1995_first}, SDSS spectroscopy has enabled unique insights into the nature and evolution of accretion in radio AGN \citep[e.g.][]{mclure2004, best2005, best2014, kauffmann2008}.
In parallel, when combined with equivalent \HI surveys such as the Arecibo Legacy Fast ALFA Survey \citep[ALFALFA;][]{giovanelli2005}, the same parent spectroscopic samples have made it possible to constrain fundamental scaling relations and extend \HI population constraints beyond the limits of blind source catalogues \citep{brown2015_alfalfa, maddox2015_sdss}.

In the decade following the publication of the first edition of Advancing Astrophysics with the SKA \citep[AASKA;][]{bourke2015advancing}, a number of planned massively multiplexed spectroscopic instruments and facilities have become a reality and at the point of this chapter's publication are either commencing, or are well into, full survey operations.
Leading this new generation is the Dark Energy Spectroscopic Instrument \citep[DESI;][]{desi_overview} that is observing homogeneously selected galaxy and QSO samples over the northern hemisphere that will be an order of magnitude larger than equivalent SDSS samples \citep{desi_validation}.
Even when combined with the same legacy radio surveys, the enormous wealth of new optical spectroscopy is extending observational constraints to new regimes \citep[e.g.][with ALFALFA]{scholte2024_desi}.
DESI will be joined in the Northern hemisphere by the WHT Enhanced Area Velocity Explorer \citep[WEAVE;][]{jin2024_weave}; while not expected to reach the same enormous sample sizes of DESI, the WEAVE Survey programme will include the first systematic fibre spectroscopy of large samples of purely radio continuum selected sources \citep[$\lesssim10^6$;][]{smith2016_wl}.

Most relevant to future SKAO surveys, new spectroscopic facilities in the Southern hemisphere have also advanced from proposals to physical instruments and detailed survey programmes.
The 4-metre Multi-Object Spectroscopic Telescope \citep[4MOST;][]{dejong2019_4most} and the first generation of 5-year 4MOST surveys will not only extend SDSS-like spectroscopic samples to the Southern hemisphere \citep{taylor2023_4hs} but provide large and highly complete samples of galaxies \citep{driver2019_waves}, X-ray/infrared/variability selected AGN \citep{merloni2019_agn, Bauer2023} and radio continuum selected sources from SKAO pathfinders \citep{duncan2023_orchidss}.
4MOST will be complemented by the Multi-Object Optical and Near-infrared Spectrograph \citep[MOONS;][]{cirasuolo2020_moons} on the 8-metre Very Large Telescope (VLT), extending large-scale galaxy spectroscopy surveys to new wavelength (and hence redshift) regimes \citep{maiolino2020_moonrise}.

With this extraordinary wealth of new spectroscopic survey facilities operational several years in advance of full SKAO operations, the community is well placed to lay the groundwork of complementary spectroscopic datasets that will enable immediate and rapid exploitation of SKAO surveys.
In this chapter we outline the scientific potential of combining SKAO radio continuum and \HI observations with these unprecedented samples of optical to near-infrared fibre spectra.
Initially, we explore the synergies between early SKAO observations (with both AA$^\star$ and AA4) and the full suite of ongoing or planned surveys that will be in place at this time.
In addition, we then set key considerations for future fibre spectroscopy surveys that maximise the potential synergies with full SKAO AA4 surveys.

Throughout this chapter, we will focus specifically on the synergies with fibre-fed multi-object spectrographs and associated surveys with total sample sizes of order $N\sim10^5 - 10^6$.
We therefore do not include discussion of advances in \emph{resolved} optical spectroscopy, for which unique synergies with radio continuum/\HI \citep[][]{masters2019_mangahi} also exist, but are typically limited to smaller and more targeted spectroscopic samples as well as typically lower redshifts.
Similarly, given the large fields of view and survey efficiency of SKAO and its pre-cursors and pathfinders, we focus on spectroscopic survey facilities that are well matched to the minimum areas for deep and wide radio surveys ($\gtrsim1\,\text{deg}^{2}$).
However, we note that many of the key scientific diagnostics enabled by the combination of SKAO with optical spectroscopy discussed here will nevertheless also extend to both resolved or detailed follow-up observations of smaller samples.
Finally, we also omit discussion of the science enabled by blind slit-less spectroscopy from current \citep[\emph{Euclid};][]{euclid2025_overview} or future \citep[\emph{Roman}; see e.g.][]{roman2025} space-based survey facilities \citep[see also][]{Prandoni01.2026.SKA}. 

Throughout this chapter, all magnitudes are quoted in the AB system \citep{OkeGunn1983} unless otherwise stated.
We also assume a $\Lambda$ Cold Dark Matter ($\Lambda$CDM) cosmology with $H_{0} = 70$ km\,s$^{-1}$\,Mpc$^{-1}$, $\Omega_{m}=0.3$ and $\Omega_{\Lambda}=0.7$.


\section{The power of optical spectroscopy for radio continuum and \HI samples}\label{sec:spectroscopy}
The richness and complexity of astrophysical and cosmological information provided by large-scale spectroscopic surveys such as SDSS is widely established.
However, to provide context and motivation for the specific synergies enabled by combining these datasets with future SKAO observations it is valuable to summarise the key astrophysical quantities probed by these spectroscopic surveys for the radio-selected galaxy and AGN populations. 

\begin{figure}[H]
    \centering
    \vspace{-0.5cm}
	\includegraphics[width=1\columnwidth]{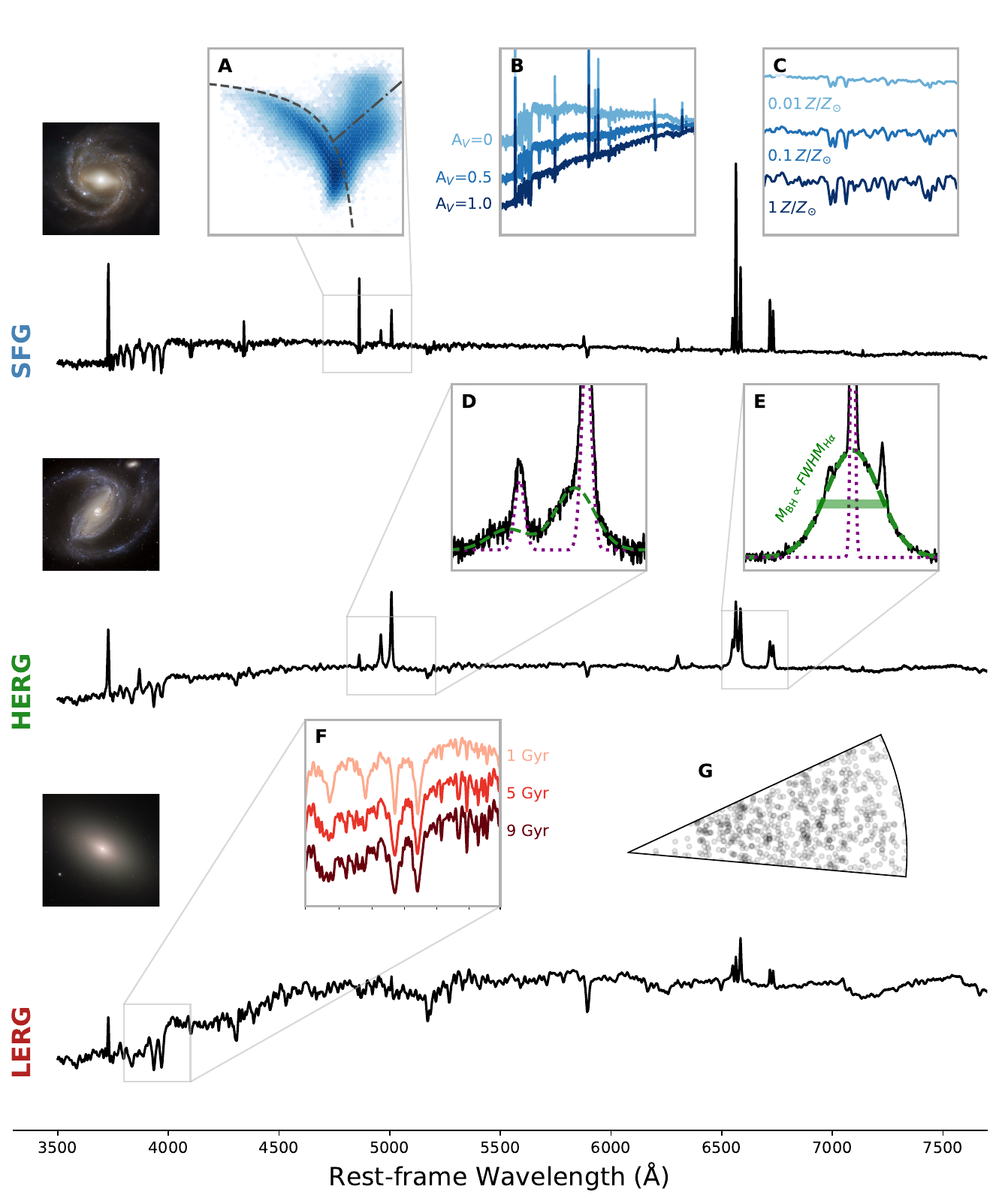}
    \caption{Visual summary of key observables unlocked by optical spectroscopy. We show stacked composite spectra of radio continuum detected galaxies classified as star-forming galaxies (SFG), high-excitation radio galaxies (HERGs) and low-excitation radio galaxies (LERGs) following the approach presented in \citet{arnaudova2025}. Inset panels {\bf A - G} are discussed in further detail in the main body of the text -- we note, however, that most of the probes featured in the inset axes are potentially applicable to all three galaxy types. The thumbnail galaxy images are chosen to be nearby galaxies that are loosely illustrative of the corresponding composite spectra: NGC 7773 as a barred spiral star-forming galaxy, NGC 1097 as a star-forming galaxy with a Seyfert nucleus, Messier 59 as a luminous passive galaxy typical of LERG hosts.}
    \label{fig:spec_overview}
\end{figure}

\subsection{Redshifts}\label{sec:redshifts}
Firstly, the most fundamental quantity provided by optical spectroscopy is precise measurements of galaxy redshifts (i.e. spectroscopic redshifts or spec-$z$s). 
While many subsets of the radio continuum selected population are well suited to measurements of precise and robust photometric redshifts (photo-$z$s; i.e. star-forming galaxies, or massive passive galaxies), the enormous diversity of AGN populations probed by radio continuum selection can make photo-$z$ estimates for these samples significantly more difficult than purely optically selected samples \citep[e.g.][]{duncan2018_photoz}.
Effort into developing appropriate observational templates \citep{brown2019_agn} and novel applications of machine learning techniques \citep{hatfield2020_gpz, duncan2022_photoz} have enabled a broad range of science for SKAO pre-cursor and pathfinder surveys \citep[see e.g.][and works which reference these]{duncan2021_dfdr1, hatfield2022_hyb}.
However, even for the very best photo-$z$s, the uncertainties on individual redshifts can for example significantly impact the key astrophysical properties inferred from spectral energy distribution modelling \citep{battisti2019} and dilute or bias measurements of local/global environments \citep{Cooper2005}.
Therefore, while photo-$z$s will likely provide the redshift estimates for the majority of sources without blind \HI detections and still enable extraordinary science, high precision redshifts for large statistical and representative samples will be essential for achieving many broader SKAO science goals. 


As will be described in Section~\ref{sec:synergies-aas-hi}, \HI\ stacking experiments require the redshift precision afforded by sufficiently high resolution spectroscopy. However, the potential scientific synergies from combining SKAO \HI observations with new spec-$z$ samples extend beyond just enabling ensemble statistical studies.


With a precise spectroscopic redshift in hand, the available suite of information that can be extracted from an optical spectrum (or sample thereof) is highly dependent on not just the properties of the spectrum (signal-to-noise, wavelength coverage and spectral resolution), but also crucially the nature of the source itself. 
Outlining the full suite of possible properties and their respective limitations/caveats are beyond the scope of this chapter.
In Fig.~\ref{fig:spec_overview} and the corresponding text below, we therefore attempt only to summarise and illustrate the most relevant properties for different subsets of the radio continuum and \HI selected galaxy populations.

\subsection{Emission line properties}\label{sec:emission_lines}

\textbf{Source classifications:}\, Both for radio continuum selected samples (which preferentially selects on activity, either star-formation or AGN) and optically selected samples for other scientific goals, robust source classifications are essential for understanding the demographics of the populations of interest and for enabling the curation of clean samples of the desired galaxy populations for further study.

Here, emission line diagnostics provide powerful tools for characterising the nature of the source population. 
Standard emission line diagnostic diagrams such as the Baldwin-Phillips-Terlevich \citep[BPT;][]{baldwin1981_bpt, veilleux1987, cidfernanders2011} diagram (Fig.~\ref{fig:spec_overview} panel A) and mass--excitation \citep[MEx;][]{juneau2011} classification enable robust separation of star-forming galaxies from AGN-dominated systems, which is essential for identifying the dominant emission mechanisms responsible for radio continuum emission.
Furthermore, for sources with clear radio continuum excesses, such classification schemes are able to distinguish between likely accretion modes through the separation of high-excitation radio galaxies (HERGs) and low-excitation radio galaxies \citep[LERGs:][]{best2012}.
In Fig.~\ref{fig:spec_overview}, we present composite spectra of radio continuum sources that are robustly identified as each of three broad classifications (SFG/HERG/LERG) based on their DESI spectroscopy \citep[][radio-quiet AGN are omitted for space and due to their potential overlap in optical properties with HERGs]{arnaudova2025}.

As these standard line diagnostics are designed to separate sources with different ionisation mechanisms, they are also essential for: i) identifying emission lines associated with high velocity shocks in the ISM \citep{allen2008, flury2025}, and ii) characterising the ionisation conditions of \HII regions, including measurements of ionisation parameters, electron densities, and gas temperatures \citep[see][for a detailed review]{kewley2019}.
Both of these are particularly crucial for extracting the maximum possible information from multi-frequency radio continuum surveys with SKA-Mid by helping to disentangle the free-free radio emission associated with star-formation \citep{algera2022} and shocks from AGN-driven winds \citep{nims2015}.

\textbf{Gas-phase metallicities:}\, Metallicity and chemical abundance measurements from optical emission lines trace the chemical enrichment history of galaxies. While the most robust so-called `direct-method' calibrations are a highly valued measurement for a minority of the brightest sources (due to the requirement to detect weak auroral lines), strong-line metallicity calibrations using ratios such as \oiii/\hbeta, \nii/\halpha, $R_{23}$ and $O_{32}$ also provide measurements of gas-phase metallicity that can be compared with stellar metallicities to understand the interplay between star formation, AGN activity, and chemical evolution. 
More complete and robust measurements of gas-phase metallicity (that can only come from optical spectroscopy) are particularly valuable for SKAO \HI samples, with the role of \HI in shaping key scaling relations such as the fundamental metallicity relation \citep{mannucci2010} an important outstanding question in galaxy evolution \citep[see][and further discussion below]{scholte2024_desi}.

\textbf{Dust attenuation and star-formation rates:}
While calibrations of the radio continuum  SFR relation \citep{gurkan2018,smith2021, cook2024, das2024} and our physical understanding of mechanisms dictating the correlation \citep{heesen2024} continue to improve, dust-corrected \halpha\ luminosity remains the gold-standard estimate for measuring the SFR in galaxies \citep{kennicutt1998}.
With rest-UV-derived SFRs not accessible at $z \lesssim 2$ to sufficient depth, measurements of the observed \halpha\ luminosity and the Balmer line ratios necessary to correct for nebular dust attenuation (Fig.~\ref{fig:spec_overview} panel B) will therefore remain essential for a broad range of SKAO science goals.

\textbf{Ionised gas kinematics:}\, One of the central goals of SKAO Extragalactic Continuum surveys is to understand the role of AGN feedback in shaping galaxy evolution. 
The kinematics of ionised gas in and around galaxies probed through the observed emission line profiles will be fundamentally essential in addressing this challenge, providing crucial insights into galaxy dynamics, SMBH properties and ongoing feedback processes. 

\begin{itemize}
    \item Outflows -- Whether driven by star-formation \citep[i.e. supernova driven winds, e.g.][]{bradshaw2013, krumholz2017} or AGN activity \citep[e.g.][]{mullaney2013, harrison2014, zakamska2016}, outflows can be identified and characterised via asymmetric line profiles in both emission and absorption (Fig.~\ref{fig:spec_overview} panel D).
    When combined with measures of the electron density from density sensitive line ratios \citep[e.g. \sii, but see also][]{collins2009, holden2023}, we can estimate mass outflow rates and begin to infer the key physical quantities (e.g. kinetic power) that are required to constrain feedback models.
    
    Here, optical spectroscopy and the ionisation diagnostics provided (see above) are essential for distinguishing between wind-driven shocks (that produce free-free dominated radio emission) and jet-driven activity (that produces synchrotron-dominated emission), enabling the separation of radiative and mechanical feedback modes.
    Similarly, the line profile shapes can provide critical information on the driving mechanism for an observed outflow, distinguishing between energy driven and momentum-driven winds \citep{chisholm2016, flury2023}.
    By detecting and characterising these broad components with spectroscopy we can investigate how the observed radio continuum emission is, or is not, associated with outflows.

    Ultimately, IFU observations that can resolve the geometry and structure of the ionised gas are necessary to fully disentangle the full details of outflows and the scale of their impact on individual host galaxies \citep[e.g.][]{harrison2014}.
    However, improvements in the modelling of outflow properties from fibre spectroscopy \citep{flury2023, flury2025b} combined with the improved sensitivity and resolution of new facilities will enable constraints on samples of galaxies spanning the full dynamic range of mass, star-formation, accretion and jet activity. This information is incredibly valuable since for example understanding the role of feedback requires a holistic view of how AGN impact upon galaxies \textit{on average} and this information is best derived using massively multiplexed optical spectroscopy.

    \item Black hole masses -- Equally key in understanding the role of AGN feedback in galaxies is knowledge of the properties of the SMBH themselves.
    Reverberation mapping \citep[RM;][]{kaspi2000, peterson2004} measurements of SMBH masses, $\mbh$, are feasible for only a small subset of the AGN population.
    Single-epoch black hole mass estimates using broad-line region kinematics \citep[][see e.g. Fig.~\ref{fig:spec_overview} panel E]{mclure2002, vestergaard2006} are therefore essential for providing measurements of $\mbh$ for large statistical samples of quasars across cosmic history \citep{shen2011}.
    
    Despite their potential biases and limitations \citep[see e.g.][]{shen2008}, the combination of single epoch estimates with deep radio continuum observations can nevertheless be used to explore the role of $\mbh$ in dictating jet activity \citep{yue2025}.
    With ever deeper samples and improved monitoring, current \citep{shen2015} and future \citep{frohmaier2025_tides} RM campaigns will continue to improve our constraints on these scaling relations.
    The improved sensitivity and wavelength coverage of new spectroscopic surveys will also enable the application of these more robust single-epoch estimates to new regimes.

\end{itemize}

\subsection{Optical continuum properties}\label{sec:continuum}
The high levels of star-formation or accretion activity in radio continuum selected populations \citep{best2023}, and the corresponding higher prevalence/brightness of emission lines, mean that a wealth of information can be obtained for sources even when the optical stellar or AGN continuum is not well constrained.
However, there are many key galaxy properties for which high signal-to-noise ratio (SNR) measurements of the optical continuum are essential.

\textbf{Star-formation histories:} While dust-corrected \halpha\ luminosities quantify the recent star-formation activity in galaxies ($\lesssim 10\text{Myr}$), inferring the past activity of galaxies, or their star-formation history (SFH), requires constraints on absorption features or diagnostics such as Balmer absorption lines or $D_{4000 \textrm{\AA}}$ \citep{balogh1999} that are sensitive to older stellar populations (i.e. hundreds of Myr to several Gyr; see Fig.~\ref{fig:spec_overview} F). 
Various absorption spectral indices ($D_{4000 \textrm{\AA}}$, H\,$\delta$,  Ca \textsc{ii} H \& K) probe star formation rates at different times over the last Gyr \citep{kauffmann2003, moresco2018}. 
Modern full spectral fitting codes extend beyond simple indices, incorporating all available features \citep[see e.g.][]{tojeiro2007, carnall2019, cappellari2023} and maximising the spectral information available.
Degeneracies between age/SFH and other physical properties such as metallicity (see below) or \emph{stellar} dust attenuation can, however, limit the precision and reliability of derived constraints; it is only with moderate to high resolution ($1000 \lesssim R \lesssim 6000$) and high SNR continuum observations that the detailed fossil records of galaxies can be disentangled \citep[][with higher resolutions yielding diminishing returns due to the intrinsic velocity dispersions of more massive galaxies]{ocvirk2006, koleva2008}.

Targeted programmes in the near-infrared have extended constraints on the SFHs of galaxies (quiescent or otherwise) to cosmic noon \citep{belli2019, slob2024} and beyond \citep{schreiber2018, carnall2023}.
Nevertheless, larger statistical samples of galaxies at $z\lesssim1.5$ with robust SFH constraints are still needed to address key questions such as the detailed role of AGN feedback in quenching galaxies \citep{maiolino2020_moonrise}.
The combination of these samples with SKAO radio continuum and \HI\ observations will be critical for obtaining the corresponding census of AGN jet activity, gas content and morphology, in order to identify the mechanisms responsible for galaxy quenching.
For example, long quenching timescales occur when e.g. starvation processes, are at work, as opposed to rapid and efficient quenching via, e.g. ram-pressure stripping.

\textbf{Stellar metallicity:} Inherently linked to the SFH of galaxies are their chemical enrichment histories \citep{maiolino2019}.
Although strongly correlated with the gas-phase metallicity probed through emission line diagnostics, the metallicities of the stellar populations themselves provide a different window in the chemical evolution of galaxies (Fig.~\ref{fig:spec_overview} panel C).
Photometric and low-resolution spectroscopy observations however suffer from significant degeneracies between age, metallicity and dust \citep{gallazzi2005}.
Moderate to high resolution spectroscopy ($R\sim5000$) is therefore essential for breaking some of these degeneracies and enabling robust constraints on both the stellar population ages and metallicities.

\textbf{Stellar kinematics:} Constraints on the line of sight velocity dispersion (LOSVD) are not only essential for accurately measuring the detailed stellar population ages and metallicities outlined above \citep{koleva2008}, but can also provide critical galaxy or physical properties through scaling relations such as the Fundamental Plane \citep[FP;][]{djorgovski1987, dressler1987}.
This is particularly important for the derivation of peculiar velocity estimates and the subsequent constraints on late-stage cosmology (beyond the scope of this Chapter) or calibrations for the extension of baryonic Tully-Fisher relation measurements \citep{Ponomareva2021} with the SKAO.

\subsection{Environmental properties}
As noted in Section~\ref{sec:redshifts}, although photo-$z$ estimates of the radio continuum samples are sufficient to enable reliable estimates of some intrinsic properties (e.g. luminosity, physical size), the velocity precision offered by spec-$z$s is essential for reliably measuring the cosmological context of the population of interest across a broad range of physical scales (Fig.~\ref{fig:spec_overview} panel G).

\textbf{Large-scale structure:} On the largest scales, uniformly selected spec-$z$ samples with well characterised selection functions enable measurements of the two- and three-dimensional clustering properties of galaxies and AGN that can begin to place populations into their wider cosmological context \citep{coil2009, hickox2009}.
The accuracy offered by spec-$z$s avoids the necessity of averaging or projecting the clustering signals over wide redshift bins.
Through approaches such as halo occupation distribution (HOD) modelling \citep{berlind2002_hod}, we can then estimate minimum halo masses as a function of galaxy/AGN property \citep[e.g.][]{petter2024}, a key step in enabling comparison between observations and simulations. 
With the increased sensitivity of SKAO radio continuum surveys over wide areas likely to yield samples \textbf{$N \times$} greater than the current state of the art, clustering and HOD studies will potentially be able to sample relevant parameters (e.g. stellar/black hole mass, radio power, Eddington ratio etc.) with both improved fidelity and greater precision. 

\textbf{Galaxy clusters and groups:} On smaller scales, measuring galaxy structures such as clusters, groups, filaments and voids robustly requires highly complete spectroscopic surveys over cosmological volumes sufficiently large to probe the required dynamic range \citep{robotham2011_gama}.
Such measurements are however critical to place observational constraints on the impact of local and large-scale environments on galaxy and AGN evolution.  
Additionally, galaxies can be associated with specific host halo masses and not just ensemble averages.

\textbf{Merger identification:} 
On even more local environmental scales, galaxy mergers are known to play an important role in the mass build-up and structural evolution of galaxies since at least $z\sim1$ \citep{conselice2014}.
However, the role of galaxy mergers in triggering AGN activity remains a topic of open debate \citep{ellison2019}.
Similarly, studies of the impact of mergers on the \HI properties of galaxies have to-date been limited to small samples in the local Universe \citep{ellison2015_hi, zuo2018_himergers}.

While new generations of deep photometric surveys can identify mergers based on morphological features such as tidal tails or disturbed morphologies \citep[e.g. \emph{Euclid};][]{lamarca2025_euclidmergers} over wide areas of sky with increasing completeness, the strong redshift evolution of surface brightness dimming ($\propto(1+z)^4$) will still place limits on the types of merger signature visible at higher redshifts.
Identification of close spectroscopic pairs will therefore remain one of the most robust means of identifying major and minor galaxy mergers \citep[e.g.][]{patton2002}.

\quad

The spectroscopic sample sizes, wavelength coverage, spectral resolution and signal to noise required or best-suited to measure each of the observables or physical properties outlined in the section above will vary significantly.
The exact requirements will also vary significantly depending on the nature of the specific galaxy population being studied as well as the corresponding science objectives.
No single instrument or survey is therefore able to provide the spectroscopic datasets required to unlock the full potential of the SKAO extragalactic surveys.
Thankfully, the new generation of spectroscopic facilities and their accompanying survey programmes, such as the ones described below, cover a broad range of these requirements.

\section{Next-generation spectroscopic facilities and surveys}\label{sec:specfacilities}

For the purpose of this chapter and the scope of the potential SKAO synergies presented, we focus our main discussion on highly multiplexed spectroscopic facilities and accompanying survey programmes that are currently in operation or are expected to be in full operation by the time of AA$^\star$.
The high level technical specifications of the key facilities are summarised in Table~\ref{tab:facilities}, with more extended overviews presented in the following subsections.

\begin{table}[hb]
    \footnotesize
    \centering
    \begin{tabular}{lcccccc}
    \hline
     Spectrograph &  Operations & Mirror Diameter & Field of View & $N_{\text{fibres}}$ & $\lambda_{\text{range}}$ & $R$ \\
         \hline
    DESI    & 2021 - 2026$^{1}$ & 4m  & $8\,\text{deg}^{2}$ & 5000  &  360--980nm & 2000 -- 5000\\
    4MOST   &  2026 -& 3.8m & $4.1\,\text{deg}^{2}$ & 2436 & 370--950nm  & 6500 \\
    WEAVE   & 2026 - & 4.2m  &  $3.1\,\text{deg}^{2}$ & 960/940 & 360--980nm & 4000 -- 6000 \\
    \hline
    PFS     & 2025 - & 8m & $1.25\,\text{deg}^{2}$ & 2400 & 380--1260nm  & 2500 -- 4500 \\
    MOONS   & 2026 - & 8m & $0.14\,\text{deg}^{2}$ & 1001 & 0.6--1.8$\mu$m & 4100--6600\\
    \hline

    \end{tabular}
    \caption{Key technical specifications and properties for the prime multi-object spectroscopic facilities operating up to, and likely into, SKAO AA$^\star$ operations. Note that spectral resolutions correspond to low-resolution modes with full wavelength coverage. \newline $^{1}$ Current survey programme only, however both DESI survey extensions and new second generation surveys are expected to continue beyond this date.}
    \label{tab:facilities}
\end{table}

\subsection{DESI}
The Dark Energy Spectroscopic Instrument (DESI), mounted on the 4m Mayall Telescope (Kitt Peak, Arizona) consists of 5000 fibres spread over a 3.2\,deg diameter focal plane.
With each 1.5 arcsec diameter fibre independently positionable within a 12mm patrol field ($\sim90$ arcsec radius on-sky), the array can be re-configured for a new field in under two minutes. 
The fibres feed ten identical spectrographs with three arms spanning 360-980\,nm.
Spectral resolution varies across the wavelength range, from $R\approx 2000$ at 360–555\,nm to $R\approx 5000$ at 680–980\,nm.
The combination of increased collection area, field configuration speed and spectral resolution for DESI represent a significant advance on the SDSS/BOSS spectrographs.

Commenced in 2021, the prime five year DESI surveys programme is expected to observe $\approx 4\times10^{7}$ sources over $14\,000\,\text{deg}^{2}$. 
The DESI sub-surveys encompass a range of different populations, including the bright galaxy sample \citep[$\sim10^{7}$ sources;][]{desi_bgs}, emission-line galaxies \citep[$\sim2\times10^{7}$ sources;][]{desi_elg}, luminous red galaxies \citep[$\sim4\times10^{6}$ sources;][]{desi_lrg} and quasars \citep[$\sim3\times10^{6}$ sources;][]{desi_qso}. 

Due to the latitude of Kitt Peak observatory, DESI's primary survey footprint is focused on the Northern hemisphere, with observations extending to $\delta\sim-5\,\text{deg}$ in the North Galactic Cap and $\delta\sim-15\,\text{deg}$ in the South Galactic Cap. 
Proposed extensions to the DESI survey could extend observations to lower declination with the goal of improving overlap with key Southern hemisphere surveys such as the Vera C. Rubin Observatory Legacy Survey of Space and Time \citep[LSST;][]{ivezic2019_lsst}.
Such extensions would naturally also improve the potential overlap with wide-area SKAO surveys.

\subsection{4MOST}
Mounted at the European Southern Observatory's (ESO) 4m VISTA telescope on Cerro Paranal, the 4MOST spectrograph has 2436 fibres covering a 2.6\,deg diameter ($4.1\,\text{deg}^{2}$) hexagonal focal plane \citep{dejong2019_4most}. 
Similar to DESI, the 4MOST fibres, each subtending 1.45 arcsec on-sky, can be robotically reconfigured in under 2 min, with each fibre patrolling a field of $\sim200$ arcsec. 
A key distinction between 4MOST and DESI is that 4MOST fibres feed two low-resolution spectrographs (LRS) and one high-resolution spectrograph (HRS).
The LRS spectrographs (812 fibres each) employ three arms covering 370–950\,nm with a spectral resolution $R\sim6 500$.
The HRS unit (812 fibres) delivers $R\sim20\,000$ in three simultaneous windows: 392–456\,nm, 515–572\,nm and 610–675\,nm.
Although primarily tuned for galactic chemical-abundance and radial velocity studies, the HRS fibres still offer potential synergies with extragalactic SKAO science through probes of the circum-galactic medium (CGM) along sight-lines towards optically luminous background sources \citep{peroux2023}. 

The main 4MOST surveys programme is due to start full operations in early 2026, with an initial suite of five year surveys that amount to $\sim 4\times10^{7}$ fibre hours of observations distributed over the Southern sky ($\delta < 5\,\text{deg}$).
The full 4MOST surveys programme consists of both consortium and ESO community surveys across a broad range of science goals, of which 5 and 9 respectively are defined as extragalactic\footnote{\url{https://www.4most.eu/cms/science/overview/}}.
Most of these surveys also consist of multiple sub-surveys such that the full 4MOST programme contains >130 sub-surveys.
Summarising the potential synergies of all surveys individually is therefore beyond the scope of this chapter, however, in Section~\ref{sec:surveys} below we include relevant characteristics and summary statistics for the most relevant subset.

\subsection{WEAVE}
Mounted on the 4.2 m William Herschel Telescope (WHT) at the Roque de los Muchachos Observatory in La Palma, Spain, the WHT Enhanced Area Velocity Explorer \citep[WEAVE;][]{dalton2012} is the next-generation 4m-class spectroscopic facility for the Isaac Newton Group of Telescopes (the ING; Spain, United Kingdom, and the Netherlands). 
In addition to two modes offering resolved optical spectroscopy, WEAVE's multi-object spectrograph (MOS) mode consists of science fibres on two different plates, 960 on one, 940 on the other, with each fibre subtending 1.3 arcsec on-sky and configurable within a 2 deg-diameter (3.1 $\text{deg}^{2}$) field of view. 
Differing from DESI and 4MOST, WEAVE's robotic fibre configuration is performed by a twin-gantry pick-and-place robot that configures one field plate while the telescope simultaneously observes with the other.
Individual fibres can therefore potentially observe targets within a much larger fraction of the field and availability of fibres for individual targets is not dictated by the fixed fibre pattern (as it is with e.g. 4MOST).
The trade-off compared to the fixed fibre pattern with smaller patrol fields is the longer overall configuration time, with WEAVE designed to enable complete plate reconfigurations during longer ($\sim 1$ hour) observing blocks. 

WEAVE's fibres feed a spectrograph with both low and high-resolution observing modes, with all fibres available in both modes.
In low resolution (LR), WEAVE delivers $R\sim5000$ across two arms spanning 366–959\,nm.
In high resolution (HR) provides $R\sim20\,000$ in two simultaneous windows: either 404–465\,nm or 473–545\,nm in the blue arm, and 595–685\,nm in the red arm. 
A unique aspect of WEAVE compared to DESI and 4MOST is the option for resolved spectroscopy; WEAVE can deploy 20 deployable mini integral field units (IFUs) with 37 1.3" fibres, covering a field of view $11\times12\,\text{arcsec}^{2}$, or a single large IFU (547 2.6" diameter fibres) covering a field $90\times78\,\text{arcsec}^{2}$.

The WEAVE Surveys programme \citep{jin2024_weave}, using $\sim70\%$ of the available WHT observing hours for five years, also targets a broad range of galactic and extragalactic science cases.
Most relevant for SKAO continuum science is the WEAVE-LOFAR Survey \citep{smith2016_wl} that will observe more than $10^{6}$ spectra of sources selected from the LOFAR Two-metre Sky Survey \citep[LoTSS;][]{shimwell2017, shimwell2019} and LoTSS Deep Fields \citep{best2023} over three survey tiers sampling the luminosity and redshift plane.
We note however, that given the requirement to target LOFAR selected sources, WEAVE-LOFAR's spectroscopic observations will be concentrated at declinations not optimal for SKA-Low or Mid surveys ($\delta \gtrsim 30\, \text{deg}$).
Nevertheless, the combination of WEAVE-LOFAR and LoTSS will represent one of the key benchmarks for the combination radio continuum sources and spectroscopy prior to SKAO operations.

Beyond the currently planned WEAVE Surveys programme, $\sim30\%$ of WEAVE observing time will also be available to proposals from the communities of the ING.
At the onset of SKAO observations there may therefore be significant additional samples of WEAVE MOS observations for equatorial fields in place for combined analyses.

\subsection{PFS}

Now installed on the 8.2m Subaru telescope on Mauna Kea, Hawaii, the Prime Focus Spectrograph (PFS) extends the new generation of MOS facilities to $\sim8\text{m}$ class telescopes.
With 2386 1.12" science fibres distributed in a fixed fibre pattern over a hexagonal field-of-view spanning 1.38 deg ($\sim1.25\,\text{deg}^{2}$), PFS offers a greater target density than DESI/4MOST/WEAVE.
The PFS fibres feed four spectrographs with three arms covering an extended wavelength range of 0.38-1.26$\,\mu$m with spectral resolution ranging from $R\sim2500$ in the blue arm to $R\sim4500$ in the near-infrared arm (the central red arm can also operate in a medium resolution mode $R\sim5500$).
The high target density and extended wavelength coverage in the near-infrared of PFS, combined with Subaru's increased collecting area offer unique capabilities for deeper extragalactic surveys that complement those from the 4m facilities outlined above.

While primarily operating as an instrument available for principle investigator proposals, PFS will also perform a number of surveys as part of a coordinated ``Strategic Surveys Program" (SSP).
The Galaxy Evolution \citep{greene2022_pfsgal} and Cosmology \citep{takada2014_pfscos} SSPs draw targets from deep photometric imaging from previous Subaru Hyper-Suprime Cam SSP observations \citep{Aihara2022} that include key equatorial survey fields accessible to future SKA-Low and Mid surveys.
For details of the target samples and corresponding science goals, we refer readers to the corresponding citations above.
Like WEAVE, in addition to the larger coordinated surveys presented above, PFS is also open to PI proposals from the Subaru scientific community.
From its location on Mauna Kea, most SKAO survey areas are observable, including key southern deep fields such as the Extended Chandra Deep Field South.
Significant additional spectroscopic samples in the coming years could therefore be obtained to further enhance synergies with the SKAO.

\subsection{MOONS}
Finally, extending the next-generation multi-object spectrographs to longer wavelength regimes (and hence higher potential redshifts) is the Multi-Object Optical
and Near-infrared Spectrograph \citep[MOONS;][]{cirasuolo2020_moons}, scheduled for installation on ESO's 8m Very Large Telescope (VLT) in Cerro Paranal.
MOONS has 1001 fibres spread over the full 25 arcminute diameter field of view ($0.14\,\text{deg}^2$) of one of VLT's Unit Telescopes, with the fibres feeding two three-arm spectrographs that cover the optical to near-infrared wavelength regime from 0.64 to 1.8$\mu$m.
In low-resolution mode, the wavelength coverage is simultaneous over this full range, with spectral resolution ranging from $R\sim4100$ in the $RI$ optical arm, to $R\sim6600$ in the reddest $H$ arm.
In high-resolution mode, MOONS achieves higher resolution in the $RI$ and $H$ arms ($R\sim9200$ over 0.76 – 0.89 $\mu$m and $R\sim19100$ over 1.52 – 1.64 $\mu$m respectively), with the central $YJ$ arm fixed at $R\sim4300$ in both modes.

Within a fixed fibre pattern, MOONS' 1.2" fibres each patrol overlapping areas with a diameter of 1.5 arcminute, achieving configuration times under 2 minutes comparable to 4MOST and DESI above.
However, unlike its optical spectrograph counterparts where a small subset of fibres ($<10\%$) can be used to map the distribution of sky background with sufficient precision and reliability for scientific analysis, the increased impact of strong OH sky lines in the near-infrared means that MOONS will likely have to survey with a more conservative observing strategy (e.g. sky fibres matched to each source, or nodding from target to background, see \citealt{cirasuolo2020_moons}).
Nevertheless, even with 50\% fibre (or exposure time) efficiency, MOONS will offer transformational survey efficiency in the near-infrared and enormous complementarity to 4MOST over the Southern hemisphere.

Within the MOONS Guaranteed Time Observations programme, the planned MOONRISE extragalactic survey \citep{maiolino2020_moonrise} is designed to exploit the wavelength coverage and multiplexing of MOONS to extend SDSS-like galaxy surveys to cosmic noon.
Depending on the final observing strategy, MOONRISE will observe between $2--4.6\times10^5$ galaxies at $0.9 < z < 2.6$ across 4--7 deg$^{2}$ of the best deep extragalactic survey fields (COSMOS, XMM-LSS, eCDFS), with high-redshift samples. 
Crucially, these fields are not only targets for deep 4MOST surveys at lower redshifts, but also the fields with the deepest SKAO pathfinder continuum and \HI observations from the MIGHTEE \citep{Jarvis2016} and LADUMA \citep{Blyth2016} surveys. Combined with deep photometric observations from \emph{Euclid} and Rubin/LSST, they are therefore the most suitable targets for deep SKAO surveys.

\bgroup
\def\arraystretch{1.05}%
\begin{table}[h]
    \footnotesize
    \centering
    \begin{tabular}{p{0.36\linewidth} p{0.13\linewidth} p{0.1\linewidth} p{0.29\linewidth} }
    \hline
       Survey & Area ($\text{deg}^2$) &$N_{\text{targ}}$ ($10^{3}$) &  Selection  \\
    \hline
        \multicolumn{4}{c}{\emph{4MOST Consortium Surveys}} \\
    \hline
        Galaxy Evolution Survey\newline\citep[WAVES;][]{driver2019_waves} & 1\,170 (Wide) \newline 66 (Deep) & 880 \newline 730  &  $Z < 21.25$ \& $z_{\text{phot}} < 0.2$ \newline $Z < 21.25$ \& $z_{\text{phot}} < 0.8$ \\
        AGN Survey \citep{merloni2019_agn} & 10\,000 & 1\,000  & $r < 22.8$ \& X-ray / IR \\
        Cosmology Redshift Survey \citep[CRS;][]{Richard2019,  verdier2025} & 5\,700 (BGS) \newline7\,000 (LRG)  &  1\,300  \newline 2\,600  &  $r < 20$ \& $grz$ colours \newline $17 < z_{\text{fib}} < 21.6$ \& $grzW1$ colours\\
        Galaxy Clusters Survey \citep{finoguenov2019} & 10\,000 & 1\,700 & eROSITA Clusters (+ Filaments)\\
    \hline
    \multicolumn{4}{c}{\emph{4MOST Community Surveys}} \\
    \hline
    4MOST Complete Calibration of the Colour-Redshift Relation \citep[4C3R2;][]{Gruen2023} & 1\,170 \newline 66 & $\sim220$ \newline $~25$ & $Z < 21.5$ \& $0.2 < z < 1.55$ \newline $Z < 21.5$ \& $0.8 < z < 1.55$ \\
    4MOST Hemisphere Survey\newline\citep[4HS;][]{taylor2023_4hs} & 18\,000 & 4\,000 & $J_{\text{AB}} < 18$ \& $J-K < 0.45$ \\
    Chilean AGN/Galaxy Evolution Survey \citep[ChANGES;][]{Bauer2023} & 18\,000 \newline 18\,000 & 900 \newline 630 & $r\lesssim 21$ Variability selected AGN \newline $r\lesssim 22.5$ SED selected AGN \\
    CHileAN Cluster galaxy Evolution Survey \citep[CHANCES;][]{haines2023} & 1330 & 300 & $z < 0.07$, $r < 20.5 $ \newline $0.07 < z < 0.45$, $r < 20.5$ \\
    Optical, Radio Continuum and \HI Deep Spectroscopic Survey \citep[ORCHIDSS;][]{duncan2023_orchidss} & 12 \newline 50 & 80 \newline 80 & $S_{\nu, 1.3 \text{GHz}} > 15\mu\text{Jy}$ \& $z_{\text{phot}} < 1.4$ \newline $S_{\nu, 1.3 \text{GHz}} > 15\mu\text{Jy}$ (or colour-based selection) \&  $z_{\text{phot}} < 0.57$  \\
    \hline
    \multicolumn{4}{c}{\emph{DESI Surveys}} \\
    \hline
    Bright Galaxy Survey \citep{desi_bgs} & 14\,000 & 10\,000 & $r < 19.5 \,(20.175)$ \& $r_{\text{fib}} < 22.9$  \\
    Luminous Red Galaxy Survey \citep{desi_lrg} & 14\,000 & 8\,000 & $z_{\text{fib}} < 21.6$ \& $grzW1$ colours \\
    Emission-line Galaxy Survey \citep{desi_elg} & 14\,000 & 16\,000 & $g > 20$ \& $g_{\text{fib}} < 24.1$ \& $grz$ colours \\
    Quasars \citep{desi_qso} & 14\,000 & 2\,800 & $16.5 < r < 23$ \& random-forest-based QSO selection \\
    \hline
    \multicolumn{4}{c}{\emph{MOONS Surveys}} \\
    \hline
    MOONRISE \citep{maiolino2020_moonrise} & 7 & 460 & $H_{\text{AB}} < 23-24$, $0.9 < z_{\text{phot}} < 2.6$\\
    \hline
    \multicolumn{4}{c}{\emph{PFS Surveys}} \\
    \hline
    PFS Cosmology SSP \citep{takada2014_pfscos} & 1\,100 & 4\,000 & $grizy$, Emission line galaxies at $0.6 < z < 2.4$ \\
    PFS Galaxy SSP \citep{greene2022_pfsgal} & 14 & 366 & Multiple $i,y,J$ with photo-$z$ over $0.7 < z < 7$ \\
    \hline
    \multicolumn{4}{c}{\emph{WEAVE Surveys}} \\
    \hline
    WEAVE-LOFAR \citep{smith2016_wl} & 6500 (Wide) \newline 650-800 (Mid) \newline 50 (Deep) & 300 \newline 175-220 \newline 275 & $S_{\nu, 144 \text{MHz}} > 7$ mJy \newline $S_{\nu, 144 \text{MHz}} > 0.7$ mJy \newline $S_{\nu, 144 \text{MHz}} \gtrsim 50-100\mu\text{Jy}$  \\
    WEAVE-QSO \citep{pieri2016} & 6500 & 330 & $r < 21.5-23.5$, $z > 2.2$ \\
    \hline
    \end{tabular}
    \caption{Summary statistics for some of the key ongoing and planned major spectroscopic surveys with next generation facilities. Note that for brevity, some sub-surveys may have been merged to provide overall statistics. Final selection properties and corresponding sample sizes may remain subject to change based on commissioning information or other strategic decisions.}
    \label{tab:surveys}
\end{table}
\egroup

\begin{figure}[h!]
    \centering
	\includegraphics[width=1.0\columnwidth]{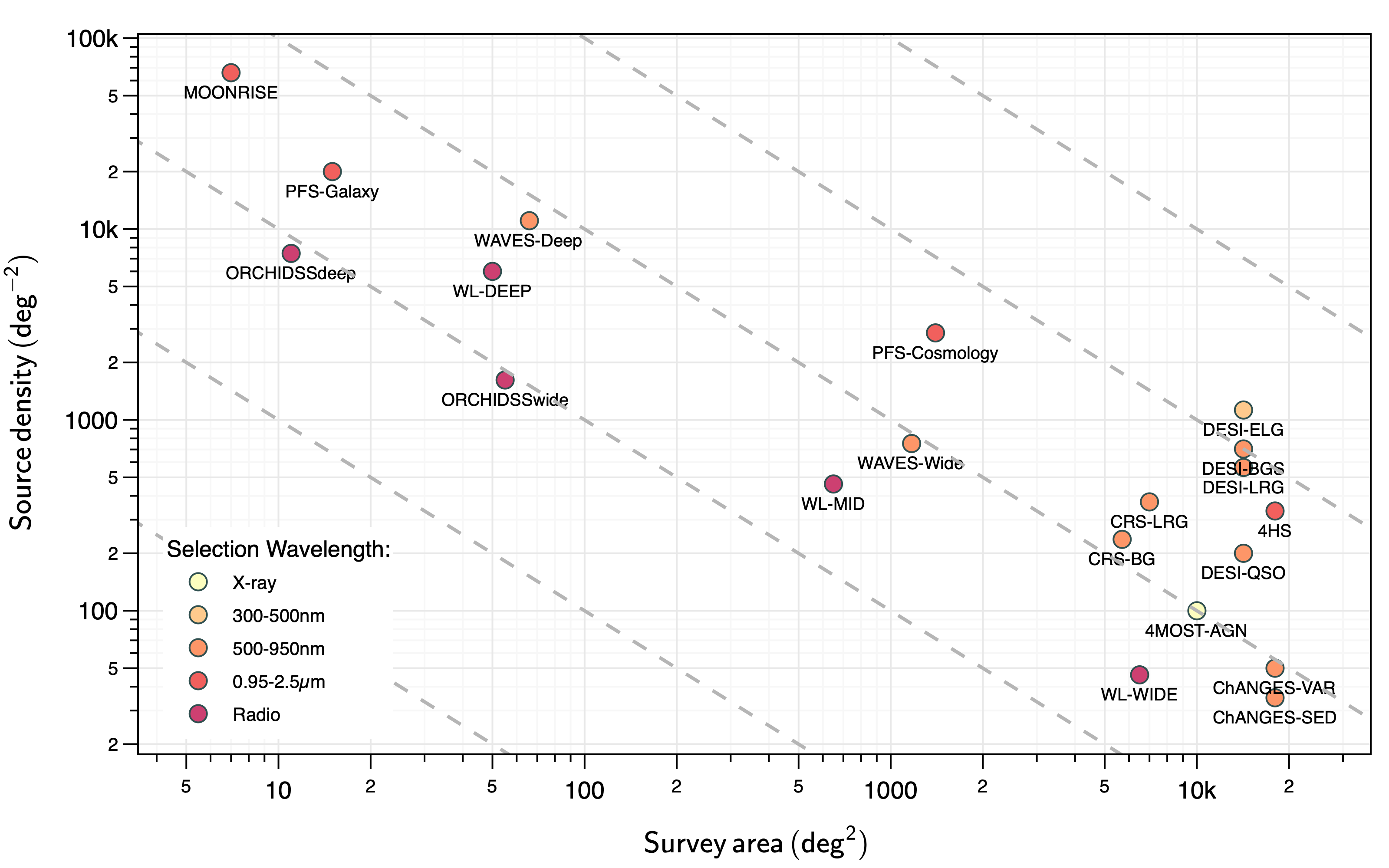}
    \vspace{-0.3cm}
    \caption{Graphical illustration of the dynamic range in survey properties of some of the current and planned spectroscopic surveys from 4MOST, DESI, MOONS, PFS and WEAVE. Figure generated directly from the \href{https://specsurveysdb.onrender.com}{Spectroscopic Surveys Dashboard}, selecting surveys with $N_{\text{gal}} > 5\times10^{5}$ and spectral resolution $R>2000$. As per Table~\ref{tab:surveys}, some survey programmes have been merged for simplicity, for full details we refer readers to the corresponding citations. Diagonal dashed lines illustrate constant samples sizes from $N=10^{3}-10^{8}$.}
    \label{fig:surveys}
\end{figure}

\subsection{Overview of key surveys}\label{sec:surveys}
Between the five facilities outlined above, the ongoing or planned spectroscopic surveys cover an enormous range of scientific goals with a corresponding diversity in sample selection, survey areas and target densities etc.
When combined with the existing wealth of spectroscopic surveys from the preceding decades, identifying the right surveys to support a given science case can become challenging.
Similarly, for this chapter, appropriately acknowledging and summarising the full range of spectroscopic surveys suitable for SKAO synergies is non-trivial.
To put the full range of available surveys into context and make them more easily searchable by relevant technical properties (e.g. spectral resolution, survey area), we have therefore constructed the \href{https://specsurveysdb.onrender.com}{Spectroscopic Surveys Dashboard}\footnote{https://specsurveysdb.onrender.com}, an interactive graphical resource.

In addition to allowing interactive plots of surveys as a function of survey area and source density, surveys can be filtered by telescope/instrument, total sample size, spectral resolution and survey status (i.e. completed/proposed).
Interactive tooltips provide additional information, as well as direct access to the best available citation DOI to make attribution and data access straight-forward and the filtered dataset can be downloaded.
Crucially, the underlying dataset can be directly updated or extended with additional or missing surveys by the community through GitHub.

In Fig.~\ref{fig:surveys}, we present an example figure generated directly from the dashboard tailored for the scope of this chapter, illustrating the range of survey areas and source densities for medium to high resolution ($R>2000$) surveys with total sample sizes above 50\,000 sources.
Collectively, the surveys included represent those from the next generation facilities most relevant to future SKAO synergies, with further details and corresponding citations of these surveys, and others, also presented in Table~\ref{tab:surveys}.

\section{Synergies between SKAO AA$^\star$/AA4 and current spectroscopic surveys}
As summarised in Section~\ref{sec:spectroscopy}, the full range of potential synergies between the diversity of ongoing/planned spectroscopic surveys and SKAO surveys is enormous and beyond the scope of a single chapter to present. 
In the following section we therefore highlight a smaller number of illustrative science cases where the combination of AA$^\star$, and particularly, AA4 with spectroscopic samples from the guaranteed surveys outlined above (particularly those from 4MOST: ORCHIDSS, WAVES, 4HS) offers immediate and significant impact on the effective sensitivity or scientific return from the radio observations.

\subsection{\HI\ in galaxies with increasing redshift}\label{sec:synergies-aas-hi}

Due to sensitivity and instrumental limitations of existing radio facilities, much of what we know about the neutral gas contents of galaxies has only been obtained from \emph{direct} \HI detections at low redshifts (e.g. \citealt{Walter2008}; \citealt{Haynes2018}). Conversely, at higher redshifts, we have been previously limited to \HI in absorption along lines of sight toward large numbers of background AGN (e.g. \citealt{Noterdaeme2012}; \citealt{Rao2017}; \citealt{Bird2017}). The redshift range $0.1<z<1$, corresponding to billions of years of cosmic lookback time, remains largely unexplored in emission.

A number of programmes designed specifically to directly detect galaxies in \HI at increasing redshifts, including the Blind Ultra Deep HI Environmental Survey (BUDHIES; \citealt{Verheijen2007}), the COSMOS \HI Large Extragalactic Survey (CHILES; \citealt{Luber2025}), and the FAST Ultra-Deep Survey (\citealt{Xi2024}) have succeeded in detecting small samples of galaxies over $0.1<z<0.4$. These precious detections come from expensive investments of telescope time and computing resources, and are thus unfeasible to extend to wider areas or larger samples. Statistical stacking techniques can alleviate some of the sensitivity restrictions, at the expense of detailed information on individual galaxies. Scaling relations for the \HI properties of galaxies have been determined at $z\sim0.4$ (e.g. \citealt{Sinigaglia2022}) and beyond $z>1$ (e.g. \citealt{Chowdhury2024}). With the vastly increased sensitivity and fields of view of next-generation radio facilities, this previously poorly probed redshift parameter space is becoming accessible with manageable observing times. The situation improves even further with SKAO AA$^{\star}$ and AA4, coupled with multiplexed optical spectroscopy, bringing the $z>0.5$ \HI universe into view, and thus bridging the gap between emission and absorption observations at a critical epoch of galaxy evolution. We outline below the galaxy samples that can be assembled from modest telescope time investment, assisted by ancillary optical spectroscopy, along with the scientific questions that can be addressed. We focus here on direct detections at  $0.23<z<0.5$, corresponding to frequencies for the \HI spectral line at 1420MHz of 950--1150MHz, within reach of both the MeerKAT L-band, and the SKA-Mid Band 2 receiver. We extend to higher redshifts and lower masses via spectral stacking, and the lower frequency SKA-Mid Band 1 receiver. 

\subsubsection{Directly detecting \HI\ in galaxies}

A number of large \HI\ programmes with SKAO precursor and pathfinder telescopes are currently underway, including MIGHTEE \citep{Jarvis2016} and LADUMA \citep{Blyth2016} with MeerKAT, and WALLABY \citep{Koribalski2020} and DINGO \citep{Rhee2023} with ASKAP, probing different combinations of area and depth, but all focusing on \HI\ beyond the local universe in large samples of galaxies. The \HI\ component of MIGHTEE \citep{Maddox2021} has generated a number of results that have already extended our knowledge of the \HI\ content of galaxies to $z\sim 0.1$ (e.g. \citealt{Ponomareva2021}; \citealt{Rajohnson2022}) and beyond (e.g. \citealt{Jarvis2024}). The first data release of spectral line cubes (DR1; \citealt{Heywood2024}), covering $\sim$4 square degrees in the COSMOS field, is a useful proxy for SKA-era datasets. It provides a baseline for comparison, illustrating the power of the SKAO. The COSMOS field also has a wealth of multi-wavelength imaging and extensive optical spectroscopy, representative of the level of completeness the upcoming spectroscopic campaigns described in Section~\ref{sec:specfacilities} are aiming to achieve, highlighting how the datasets can work together.

\begin{figure}[h]
    \centering
	\includegraphics[width=0.9\columnwidth]{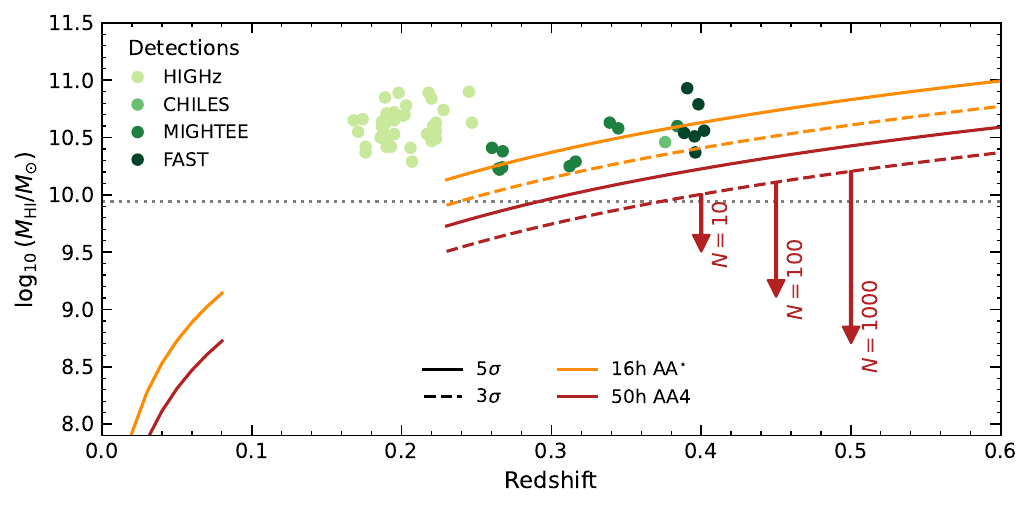}
    \caption{Illustrative redshift parameter space quickly accessible to AA$^{\star}$ and AA4, along with a compilation of existing detections at these redshifts (filled circles, see text for citations). The limiting \HI\ mass observable as a function of redshift is shown for two baseline surveys. The orange and red lines show the predictions for the 16\,h AA$^{\star}$ and 50\,h AA4  surveys respectively, with solid lines restricted to $5\sigma$ and dashed lines relaxing the detection limit to $3\sigma$. Annotated arrows illustrate the increased depth obtainable from stacking bins of 10--1000 sources (e.g. as a function of stellar mass, SFR, environment etc.). The grey dotted line indicates the characteristic \HI\ mass, $\phi$*, at $z=0$. }
    \label{fig:MinMHI}
    \vspace{-0.3cm}
\end{figure}

The MIGHTEE COSMOS area is a closely-spaced mosaic of 15 MeerKAT pointings, with effective on-source integration time per pixel of approximately 70 hours from a total observing time of 119.9\,h. The depth is sufficient to directly detect the most \HI-massive galaxies at $z\sim 0.3$ at high significance \citep[][see also Fig.~\ref{fig:MinMHI}]{Jarvis2025}. From the SKAO Sensitivity Calculator, the equivalent per-channel RMS will be achieved by AA$^{\star}$ in just 8 hours, opening the opportunity to cover substantially more area in the same amount of time.

Employing basic simulations based on those in \cite{Maddox2016}, which couple an input $z=0$ \HI\ mass function (HIMF) with given RMS flux levels, we can estimate the number of galaxies we expect to detect for a given set of observing parameters, per square degree.  For untargeted \HI\ detections, we impose a $5\sigma$ requirement on detections, to ensure a statistically robust sample. We consider two baseline observation types, to illustrate the parameter space that can be probed with each. We first consider a 16\,h integration with the AA$^{\star}$ array, which would be suitable for covering an area of $\sim$50 deg$^{2}$ or larger in a modest total time. We also consider deeper, 50\,h, integrations with the full AA4 configuration covering 10 deg$^{2}$. We exclude \HI\ detections in the redshift range $0.1<z<0.23$, which is heavily affected by radio frequency interference (RFI). For reference, an angular size of 1 degree, comparable to the size of the SKA-Mid field of view, corresponds to a linear size of 19 Mpc at $z=0.4$. Therefore, fields covering $\sim$10's of square degrees are sufficient to encompass large scale structure, including filaments, voids, and galaxy clusters. 

Fig.~\ref{fig:MinMHI} shows the approximate \HI\ mass limits as a function of redshift for these two baseline observations, 16h with AA$^{\star}$ and 50\,h with AA4. Of most interest are the mass limits above $z>0.23$. The solid lines show the $5\sigma$ limits, assuming galaxies of width $200\kms$ and 108 kHz ($31\kms$ at $z=0$) channel widths. The grey dashed line indicates the $z=0$ value for the characteristic mass, $M_{\HI}^{\star}$. Also included are existing direct \HI detections from \cite{Catinella2015} at $z\sim0.2$, \cite{Jarvis2025} at $z\sim0.3$, \cite{Fernandez2016} at $z=0.376$, and \cite{Xi2024} at $z\sim0.4$, illustrating the dearth of data available at these redshifts. The large collecting area afforded by FAST, able to detect galaxies to $z\sim0.4$, is balanced by the lack of resolution from the single dish. With moderate integrations with AA$^{\star}$, we can populate the parameter space of \HI-massive galaxies down to the characteristic mass at $z\sim0.3$ with spatially resolved detections. 

Only with the increased sensitivity of AA4 does it become feasible to probe individual galaxies at $\sim M_{\HI}^{\star}$ over representative volumes at $z > 0.2$ with only moderate time investments (see Fig.~\ref{fig:MinMHI}).
Furthermore, even with the sensitivity offered by AA4, targeting fields covered by the spectroscopic programmes described in Section~\ref{sec:specfacilities} allows us to extract even more information from the radio observations. Identifying galaxies likely to be \HI-rich via morphology and colour selection from optical imaging, and confirming their redshift with existing spectroscopy, allows us to locate the galaxies within the spectral cubes. Given these spectroscopic priors, \HI\ detections to lower significance, down from 5 to 3$\sigma$, can be extracted with confidence, as illustrated in \citet{Jarvis2025}. Shown as the dashed lines in Fig.~\ref{fig:MinMHI}, decreasing the detection threshold from $5\sigma$ to $3\sigma$ lowers the minimum detectable mass, and thus increases the number of galaxies detected per square degree by a factor of two to three.

\begin{figure}[h]
    \centering
	\includegraphics[width=0.9\columnwidth]{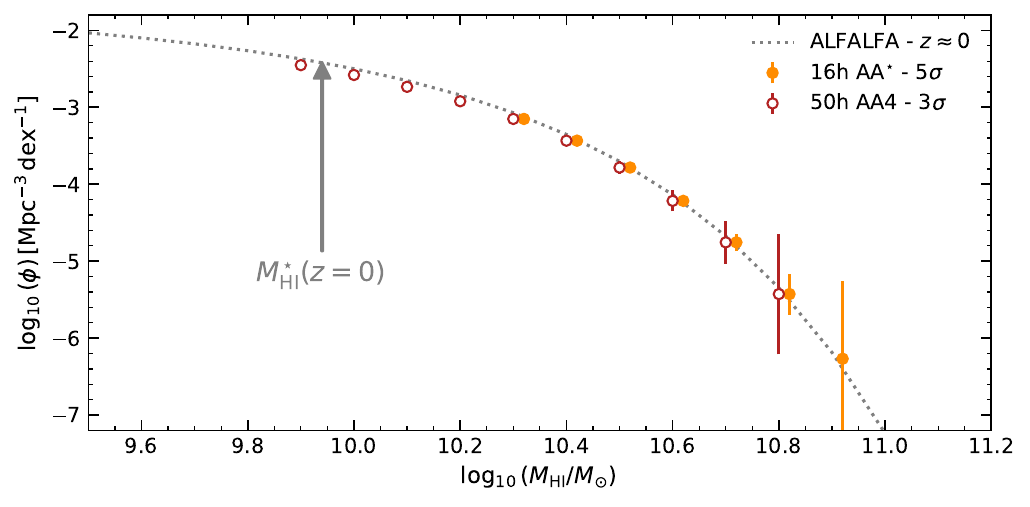}
    \caption{The parametrisation of the HIMF at $z\sim0.3$ enabled by a survey with 16\,h integrations over 50 deg$^2$ with AA$^{\star}$ (orange, $5\sigma$ detections) and 10 deg$^2$ with 50\,h integrations with AA4 (red) assuming spectroscopic prior driven detections down to $3\sigma$ significant. The $z=0$ HIMF from \citet{Jones2018} is shown in grey, with the characteristic mass, $M^{\star}_{\HI}$, highlighted.}
    \label{fig:HIMF}
    \vspace{-0.3cm}
\end{figure}

The parametrisation of the \HI\ mass function (HIMF) at $z=0.3$ enabled by the two baseline surveys is shown in Fig.~\ref{fig:HIMF}, along with the $z=0$ HIMF from \citet{Jones2018}. The depth of the surveys well parametrises the high-mass end of the HIMF, and enables the turnover mass, $M_{\HI}^{\star}$, to be probed. The simulations here do not incorporate an evolving HIMF, so any redshift evolution in high-mass galaxies will be evident from the observations. Combining these wider surveys, which probe the high-mass end of the HIMF, with deeper, smaller area \HI-focused surveys, which probe further down the HIMF but lack the volume to fully sample the high-mass end, will enable full parametrisation of the HIMF to $z>0.5$.

Not only does the addition of optical spectroscopy allow us to increase the number of galaxies that can be detected confidently, it also provides additional information for each detection. First, we can easily distinguish between the \HI\ 1420MHz and OH 1667MHz emission lines. OH megamasers have been discovered at increasing redshift (e.g. \citealt{Glowacki2022}; \citealt{Jarvis2024}) and can be a significant contaminant in \HI\ observations (\citealt{Roberts2021}). The highly complete spectroscopy also gives a robust measure of the wider environment within which the \HI-detected galaxies reside. \HI\ is known to be very sensitive to the immediate (e.g. \citealt{Jones2020}) and large-scale structure \citep{Tudorache2022}. 
As per Section~\ref{sec:emission_lines}, the metallicity of the galaxy ISM can also be determined, indicating which objects are most promising for subsequent observations of the molecular gas \citep{Tacconi2018}. 
Similarly, optical emission line diagnostics can robustly identify AGN, and their effect on the gas content of the hosting galaxies (e.g. \citealt{Ellison2019_HI}).

\subsubsection{Statistical \HI\ detections}

With AA$^{\star}$, and AA4, directly detecting \HI\ in massive galaxies at $z\sim 0.4$ is feasible over areas large enough to sample multiple environments with moderate investments of observing time. Including prior spectroscopic and photometric data over the same areas lowers the detection threshold, increasing the resulting sample sizes. To probe even further into the low-SNR regime, stacking the \HI\ datacubes on the known position and redshift of galaxies results in a statistical measure of the input galaxy ensemble.
At moderate redshifts, \citet{Bera2019}, \citet{Bera2023}, \citet{Luber2025}, \citet{Sinigaglia2022}, \citet{Bianchetti2025}, and \citet{pan2025}, among others, have extracted ensemble \HI\ detections and scaling relations. Studies such as \citet{chowdhury2020, chowdhury2022} have extended the redshift regimes for which statistical detections of \HI in galaxies can be made to $z>1$. 
While many \HI\ stacking efforts focus on reaching increasing redshifts, the addition of newly acquired spec-$z$s to existing low-redshift \HI\ observations can extend the galaxy parameter space for which ensemble \HI constraints can be made \citep{scholte2024_desi}. Each of these works share the common requirement of extensive and resource-intensive spectroscopic redshift observing campaigns, often undertaken by other groups as part of other projects. This results in complex selection functions, which in turn complicate interpretation of the results. The fraction of sky with suitably dense spectroscopic observations is currently restricted to a few well-studied extragalactic fields. The result from \citet{chowdhury2022}, for example, required more than $10^4$ spectra.

The wavelengths covered by the 4MOST spectrograph, 3700--9500 \AA, are well-suited to observations targeting the moderate redshift, $z\sim 0.4$ regime. The combination of field-of-view, sensitivity, and high multiplexing, enables 4MOST to cover large patches of sky suitable for \HI\ stacking relatively quickly. The resulting large, homogeneously selected spec-$z$ samples from approved surveys such as ORCHIDSS \citep{duncan2023_orchidss} and WAVES \citep{driver2019_waves} can be divided into bins of stellar mass, star formation rate, and environment, all of which show indications of affecting \HI content (e.g. \citealt{sinigaglia2024}). 
In Fig.~\ref{fig:MinMHI}, we illustrate the increased parameter space probed by stacking samples from 10--1000 galaxies.
We note that, assuming the nominal 50\,h AA4 survey targets the same fields, ORCHIDSS alone is expected to provide $\sim6\,000$ galaxies per $\delta z=0.1$ slice at the illustrated redshifts; the fidelity with which ensemble detections could constrain \HI scaling relations is therefore extremely high, even for naive stacking approaches \citep[with more novel ensemble modeling approaches likely to yield more robust constraints, e.g.][]{pan2023, pan2025}. 

However, at increasing redshifts, the primary emission lines illustrated in Fig.~\ref{fig:spec_overview}, including \halpha, \hbeta, \oiii\ and \oii\ move from the optical wavelengths to longer near-infrared (NIR) wavelengths. While 4MOST samples are expected to reach $z\approx1.4$ due to the visibility and ease of identification of the \oii\ doublet, the extended wavelength coverage and greater collecting area of the PFS and MOONS spectrographs described in Section~\ref{sec:specfacilities} will be critical for extending the parameter space accessible for \HI\ stacking at $z>1$ and providing the robust source classifications and galaxy properties needed to maximise their interpretation. 
With so little known about the \HI\ emission universe at $z=1$, it is difficult to predict what any given observation will uncover. It is likely that substantial integration times with SKAO AA4 in Band 1 will be required to directly detect galaxies in \HI\ at $z > 1$, even with the advantage of spectroscopic priors. Therefore, the information that can be extracted by stacking becomes even more valuable, and the input spec-$z$ samples become more important. 

\subsection{AGN and galaxy evolution continuum science}\label{sed:synergies-aas-cont}


\subsubsection{The role of environment in radio AGN}
As highlighted in the first edition of ASKAA, the combined sensitivity and survey speed of the SKAO should enable a highly complete sample of AGN across all environmental densities \citep{smolcic2015}.
The enhanced spatial resolution of AA4 will be essential for providing robust morphological classifications of the extended jet population. 
Based on the anticipated SKAO performance \citep{braun2019anticipatedperformancesquarekilometre}, the maximum SKA-Mid Band 2 resolution of 0.4 arcsec will provide sub-kpc resolution out to $z \lesssim0.15$ while still offering rapid survey speeds for large areas. 
Even at higher redshifts, this resolution will nevertheless exceed the resolution achieved by matching ground-based optical surveys (and be only marginally lower than space-based imaging from \emph{Euclid}; \citealt{euclid2025_overview}) and will be sufficient to constrain the distribution of jet sizes and morphologies over much of cosmic history \citep{sweijen2025}.
Despite this potentially dramatic improvement in the observed radio properties, understanding the detailed cosmological context of the observed radio AGN population remains of critical importance in extracting the underlying physics, including disentangling the roles of environment and host galaxy/SMBH properties on dictating jet activity and its evolution, or even simply accurately interpreting the observable properties of extended jets \citep[$>10$s of kpc;][]{hardcastle2020}.
Any future suite of surveys should therefore ideally maximise overlap with uniform and complete spectroscopic surveys over the same cosmic volumes.

If we assume that large continuum surveys (e.g. Key Science Projects) with SKAO AA4 follow the proven strategy of tiered surveys that efficiently span the dynamic range of area and sensitivity, the planned 4MOST spectroscopic surveys offer ideal targets over a range of well-matched scales:
\begin{itemize}
    \item Large area / hemisphere ($\sim10\,000\,\text{deg}^{2}$): Designed to provide an SDSS-like spectroscopic baseline of local Universe galaxies, the 4MOST Hemisphere Survey \citep[4HS;][]{taylor2023_4hs} will survey all of the Southern hemisphere extragalactic sky with a sample selection (see Table~\ref{tab:surveys}) that is designed to provide volume complete samples of $\log_{10}(M/M_{\odot}) > 10$ at $z < 0.1$. 
    In addition, the 4MOST Galaxy Clusters Survey \citep{finoguenov2019} and CHileAN Cluster galaxy Evolution Survey \citep[CHANCES;][]{haines2023} will provide highly complete spectroscopic coverage of massive clusters and their surrounding filaments and in-fall regions over much of the same cosmic volume. 
    Together, these surveys will provide the precise environmental information required to put the observed jet activity in cosmological context on the scale of both the extended cosmic web \citep[e.g.][]{jung2025} and individual cluster environments \citep[e.g. for extended head-tail sources;][]{bushi2025}.
    
    \item Medium-deep Surveys ($\sim1\,000\,\text{deg}^{2}$): Extending both to higher redshifts ($z < 0.2$) and a more complete picture of cosmic structure over representative volumes, the Wide tier of the WAVES \citep{driver2019_waves} Galaxy Evolution Survey will observe $\sim1100\,\text{deg}^{2}$ over two fields.
    By identifying dark matter halos down to $\log_{10}(M_{\text{DM}}
    /M_{\odot}) \sim 11$, as well the broader filamentary structures and voids surrounding them, WAVES Wide will enable observed jet (and obscuration free star-formation) activity to be placed directly into the corresponding halos.
    In addition to enabling environmental studies similar to those outlined above, the halo mass estimates provided by WAVES will be critical for allowing robust direct comparison between SKAO observations and hydrodynamical simulations of galaxy (and SMBH) evolution.

    \item Deep ($\sim10-50\,\text{deg}^{2}$): Similarly, the WAVES Deep tier will identify all of the massive dark matter halos ($\log_{10}(M_{\text{DM}}/M_{\odot}) \gtrsim 14$) out to redshift $z = 0.8$ over $66\,\text{deg}^{2}$.
    Although identification of individual halos will be limited to more massive group/cluster scales, WAVES Deep will nevertheless provide robust measures of the broader environment of galaxies over the bulk of cosmic history, including identification of both major and minor mergers.
\end{itemize}

\begin{figure}[h]
    \centering
	\includegraphics[width=0.9\columnwidth]{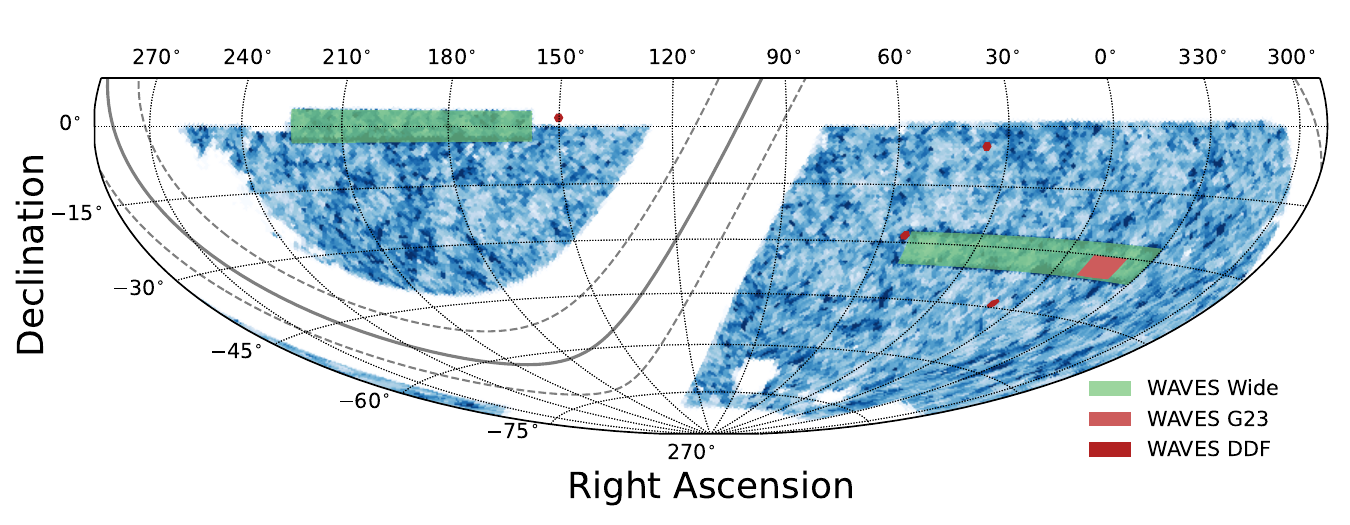}
    \caption{Illustration of key Southern hemisphere regions for which high completeness spectroscopic samples will be available from 4MOST spectroscopy (in a Hammer-Aitoff projection). The background colour scale illustrates the approximate region with complete coverage from 4HS, with the colour scale illustrating the density (arbitrary scaling) in pixels of equal volume.
    Filled regions illustrate the areas corresponding to the WAVES Wide (green) and WAVES Deep (G23 \& DDF; red) regions. }
    \label{fig:survey_fields}
\end{figure}

On the deep to medium-deep scales where specific sky regions must be selected for future surveys, targeting the spectroscopic surveys outlined above (and illustrated in Fig.~\ref{fig:survey_fields}) will offer immediate enhancement to the scientific interpretation of SKAO AA4 radio continuum samples within the corresponding redshift regimes.
As outlined in Section~\ref{sec:specfacilities} (and Table~\ref{tab:surveys}), other 4MOST surveys will provide additional spectroscopy for various galaxy and AGN subpopulations within the cosmological context provided by e.g. 4HS or WAVES.
In all cases, many of the key observables outlined in Section~\ref{sec:spectroscopy} will add further scientific value for individual targets. 
We note, however, that while these surveys will provide the broader environmental context in which to place the AGN (or \HI) population, complete samples of radio selected AGN will only be available in the ORCHIDSS fields \citep[][coinciding with the WAVES DDF fields in Fig.~\ref{fig:survey_fields}]{duncan2023_orchidss}.
Additional surveys, 4MOST or otherwise, may be required to build complete samples over the corresponding larger survey areas (see Section~\ref{sec:synergies-aa4-future} below).

\subsubsection{A detailed view of accretion and feedback at cosmic noon}
A fundamental challenge in building a complete picture of galaxy formation is understanding the role of AGN feedback in shaping the evolution of galaxies.
Observations of deep survey fields across the electromagnetic spectrum now unambiguously show that the cosmic accretion rate density peaks over the same epoch as star-formation activity \citep{aird2015}.
Spanning $1\lesssim z \lesssim 3$, this so-called `cosmic noon' therefore represents a critical period for studying feedback in action.
One of the key limitations in robustly constraining the prevalence and impact of feedback activity in this epoch has been the difficulty in obtaining the highly complete, representative samples of rest-frame optical spectroscopy necessary to provide the corresponding source classifications, ionisation diagnostics and outflow kinematics.

To-date, surveys have often been limited to small survey fields \citep[$\ll 1\text{deg}^{2}$;][]{kriek2015, wisnioski2019} or subsets of larger samples \citep{stott2016}, that contain only small or potentially biased samples of jetted AGN.
Alternatively, dedicated spectroscopic follow-up observations have typically targeted the brightest and most extreme radio sources \citep[e.g.][]{kuiper2011}.

Here, the near-infrared sensitivity and multiplexing of MOONS and the corresponding MOONRISE guaranteed time observations survey \citep{maiolino2020_moonrise} promise to revolutionise studies of galaxies in this epoch.
Targeting uniformly selected samples of galaxies in redshift bins of $0.9 < z < 1.1$, $1.2 < z < 1.7$ and $2 < z < 2.6$ (plus a higher redshift sample up to $z\sim5$), MOONRISE will provide unprecedented samples of $\sim2-4\times10^{5}$ galaxies over 4-7 deg$^{2}$ (depending on final observing strategy) that are extremely well matched to deep SKAO AA4 continuum surveys.
For the baseline deep survey assumed in this chapter, with 50\,h  on-source exposure for an area of 10\,deg$^{2}$, the nominal $5\sigma$ 1.4 GHz continuum sensitivity would reach $1.5\mu\text{Jy\,beam}^{-1}$ \citep[][assuming negligible impact from confusion noise]{braun2019anticipatedperformancesquarekilometre}.

At $z=1.7$, where MOONRISE spectroscopic samples would retain access to the full suite of key diagnostic lines required for BPT source classifications (with mass excitation and other diagnostics still applicable out to $z \lesssim 2.6$) and dust-corrected \halpha\ SFR estimates, the $5\sigma$ radio luminosity limit of $\sim4\times10^{22}\,\text{W\,Hz}^{-1}$ corresponds to SFRs of $\sim13\,\text{M}_{\odot}\,\text{yr}^{-1}$  \citep[depending on the assumed calibration, see e.g.][]{duncan2020}. This SFR limit is sufficient to probe below the typical SFR of $\log_{10}(M/M_{\odot}) \sim 10$ galaxies at this redshift \citep{speagle2014}.
In addition to providing obscuration-free SFR estimates, deep SKAO AA4 observations in concert with MOONS spectroscopy would therefore enable robust selection of radio-excess AGN down to low jet powers in individual Milky-Way-like galaxies \citep[e.g. extending approaches such as][to higher redshifts]{arnaudova2025}.

With a complete picture of the ongoing star-formation and jet activity from AA4 continuum observations, high resolution spectroscopy from MOONS (both MOONRISE and additional data obtained in the coming years) will enable exploration of the presence and properties of outflows as a function of star-formation, accretion and environment through the line profiles of ionised gas in emission (e.g. \oiii, \halpha) and neutral gas in absorption (Na \textsc{i D}).
When supplemented by SKAO constraints on the molecular gas outflows \citep[e.g.][]{wagh2024}  it will be possible to constrain the full multi-phase structure of AGN driven outflows \citep{ward2024MNRAS.533.1733W} and their impact on host galaxy properties.

Building on the simple baseline AA4 Band 2 survey outlined above, the deepest tier of potential survey programme represents an opportunity to exploit the unique frequency coverage and resolution of SKA-Mid.
By combining Band 1, 2 and 5 surveys with matched sensitivity (including multiple sub-bands) over the same area, detailed per-source constraints on the radio spectral energy distribution will enable not only constraints on the synchrotron spectral slopes (e.g. to infer jet ages), but also disentangle free-free and synchrotron contributions in unresolved or compact sources \citep{tabatabaei2025}.
Again, combination of these radio continuum constraints with matching spectroscopy will be essential for correctly interpreting the radio continuum SEDs, with detailed ionisation diagnostics helping to distinguishing free-free emission produced by wind-driven shocks from the synchrotron emission driven by jet activity.

\section{Synergies between SKAO AA4 and future spectroscopic surveys}

\subsection{New surveys on current facilities}\label{sec:synergies-aa4-current}
By design, many of the spectroscopic surveys discussed in Section~\ref{sec:specfacilities} offer enormous legacy value and the potential to address many of the galaxy and AGN evolution questions central to SKAO, as well as science questions that have yet to arise. 
Nevertheless, the increased \HI and continuum discovery space unlocked by larger observing programmes from SKAO AA4 means that dedicated future spectroscopic surveys will undoubtedly be required.
Until significant samples of Southern hemisphere surveys from 4MOST/MOONS are in hand and analysed alongside the relevant SKAO pathfinder and precursor observations, precisely defining the scope of such surveys is not practical. 

We caution, however, that the broader SKAO scientific community will need to consider and plan for such surveys significantly in advance of SKAO Cycle 1 (and certainly in advance of Key Science Projects). 
This is particularly true in the case of large coordinated survey programmes like DESI or 4MOST. 
For example, the deadline for ESO Proposals for 4MOST Community Surveys in the current 4MOST Surveys programme was December 2020, with the successful surveys only now commencing in 2026.
While the lead time for potential second generation 4MOST surveys may be shorter, the enormous complexity of efficiently integrating multiple surveys into a coherent observing strategy may nevertheless require survey strategies and target catalogues for SKAO focused surveys to be considered in the coming few years.  


\subsection{Future optical spectroscopy facilities}\label{sec:synergies-aa4-future}
On longer timescales, the SKAO will be one of the cornerstone international observatories for several decades.
The spectroscopic needs for SKAO should therefore play a central role in defining the technical and operational requirements of other future instruments and facilities, or help to shape the key surveys planned for these facilities. 

For example, the SKAO will be operational alongside proposed facilities such as the Wide-field Spectroscopic Telescope \citep[WST;][]{bacon2024_wst}.
The WST is proposed as the next major ESO optical facility following the Extremely Large Telescope, designed to address the large-scale survey needs beyond those outlined in Section~\ref{sec:specfacilities}.
The current proposed specifications combine a dedicated 12m-class telescope with up to 30\,000 fibres covering a 3.1 deg$^{2}$ field of view and simultaneous spectral coverage from 370-970nm at moderate spectral resolution $R\sim4000$.
A unique proposed feature of WST is that it would also allow simultaneous observation with a large central integral field spectrograph covering 9 arcmin$^{2}$ ($9\times$ larger than MUSE).
The combination of increased multiplexing and collecting area would offer significant advances over existing facilities for the MOS capabilities alone.
This would allow for spectroscopic analysis of existing targets (e.g. \HI samples at $z \lesssim 1$) to extend to significantly higher continuum SNR for individual targets and unlock the wealth of additional information this provides (see Section~\ref{sec:continuum}).
Additionally, extending the redshift range for which highly complete spectroscopic samples could be achieved, the combination of SKAO AA4 radio continuum selected samples and WST rest-UV spectroscopy could enable a detailed picture of the star-formation and accretion history of galaxies over the first few Gyrs of cosmic history \citep{saxena2024, gloudemans2025, ighina2025, whittam2025}.

Alongside the enormous statistical studies enabled by potential WST MOS surveys, the extraordinary contiguous field of view of the central IFS will also enable deep resolved spectroscopic observations of 10s of thousands of distant galaxies.
Combination of large area IFS studies with SKAO surveys at the full AA4 resolution will enable high-resolution observations of jet activity to be spatially linked with the corresponding outflow activity on kpc scales \citep[see e.g.][]{speranza2021}, or to link resolved maps of \HI with the corresponding stellar and ionized gas properties and kinematics \citep[e.g. extending studies such as][to high redshift]{deblok2024}.

\section{Summary}
We have presented an overview of the key physical observables and potential synergies for SKAO extragalactic science unlocked by a new generation of massively multiplexed multi-object spectrographs.
The emission line and continuum observations provided by this new suite of instruments and their associated surveys will provide precise redshifts, robust source classifications and reveal a wealth of information that will dramatically enhance the impact of early SKA Low and Mid observations (both with AA$^{\star}$ and AA4). 
Given the huge diversity of ongoing and planned spectroscopic surveys, each with its own sample selection and potential complementarity with SKAO surveys, we have presented only a broad overview of the key facilities and surveys. 
However, to aid the SKAO community in identifying the best spectroscopic resources for their science goals, we have also developed an interactive \href{https://specsurveysdb.onrender.com}{Spectroscopic Surveys Dashboard}.

To highlight specific science cases where optical spectroscopy will enable immediate benefits for SKAO analysis, we outline their potential for \HI studies by enabling statistical measures of the \HI content of galaxies out to $z=1$ and beyond through prior-driven detections and spectral stacking analyses.
In addition, we highlight two key synergies for radio continuum observations, including the comprehensive measures of environment that will be available out to $z<0.8$ and the potential probes of feedback in action at cosmic noon.
Finally, we have identified some of the additional requirements for future SKAO continuum and \HI\ focused spectroscopic surveys beyond the currently planned generation.

\section*{Acknowledgements}
KJD acknowledges support from the Science and Technology Facilities Council (STFC) through an Ernest Rutherford Fellowship (grant number ST/W003120/1).
DJBS acknowledges support from STFC via grant ST/Y001028/1, and from the Leverhulme Trust via Research Project Grant RPG-2025-078. 
MJJ  acknowledges the support of a UKRI Frontiers Research Grant [EP/X026639/1], which was selected by the European Research Council, and the STFC consolidated grants [ST/S000488/1] and [ST/W000903/1]. MJJ also acknowledges support from the Oxford Hintze Centre for Astrophysical Surveys which is funded through generous support from the Hintze Family Charitable Foundation. 











\clearpage

\bibliographystyle{abbrvnat-maxbibnames4}
\bibliography{chapter} 

\end{document}